\documentclass[a4paper, 10pt, conference, twocolumn]{IEEEtran}

\usepackage[utf8]{inputenc}

\usepackage{url}

\usepackage{breakurl}
\usepackage[breaklinks]{hyperref}
\usepackage{graphicx}
\usepackage{glossaries}
\usepackage{caption}
\usepackage{todonotes}
\usepackage{booktabs}

\newacronym{3GPP}{3GPP}{3rd Generation Partnership Project}
\newacronym{5G}{5G}{Fifth Generation}
\newacronym{API}{API}{application programming interface}
\newacronym{AR}{AR}{augmented reality}
\newacronym{CapEx}{CapEx}{capital expenditure}
\newacronym{CD}{CD}{continuous deployment}
\newacronym{CI}{CI}{continuous integration}
\newacronym{CM}{CM}{continuous monitoring}
\newacronym{DevOps}{DevOps}{Development Operations}
\newacronym{DSP}{DSP}{digital signal processor}
\newacronym{E2E}{E2E}{end-to-end}
\newacronym{EPC}{EPC}{Evolved Packet Core}
\newacronym{HSS}{HSS}{home subscriber server}
\newacronym{LTE}{LTE}{Long Term Evolution}
\newacronym{MEC}{MEC}{multi-access edge computing}
\newacronym{MNO}{MNO}{mobile network operator}
\newacronym{NFV}{NFV}{network function virtualization}
\newacronym{NIC}{NIC}{network interface controller}
\newacronym{OpEx}{OpEx}{operating expenditure}
\newacronym{OSS}{OSS}{open-source software}
\newacronym{OTA}{OTA}{over-the-air}
\newacronym{QoS}{QoS}{quality of service}
\newacronym{RAN}{RAN}{radio access network}
\newacronym{SDN}{SDN}{software defined network}
\newacronym{SIM}{SIM}{subscriber identity module}
\newacronym{UE}{UE}{user equipment}
\newacronym{SDR}{SDR}{software-defined radio}
\newacronym{PaaS}{PaaS}{Platform-as-a-Service}
\newacronym{IaaS}{IaaS}{Infrastructure-as-a-Service}

\title{Open-source RANs in practice: an over-the-air
deployment for 5G MEC}
\date{February 2019}

\author{\thanks{This is the accepted version of the work. The final version will be published at the European Conference on Networks and Communications (EuCNC2019), June 18-21, Valencia, Spain, 2019.}
\IEEEauthorblockN{Juuso Haavisto,
Muhammad Arif, 
Lauri Lov\'en,
Teemu Lepp\"anen and
Jukka Riekki}
\IEEEauthorblockA{
University of Oulu, Oulu, Finland \\
\{firstname.lastname\}@oulu.fi
}}
\IEEEoverridecommandlockouts
\begin{document}

\maketitle

\begin{abstract}

Edge computing that leverages cloud resources to the proximity of user devices is seen as the future infrastructure for distributed applications. However, developing and deploying edge applications, that rely on cellular networks, is burdensome. Such network infrastructures are often based on proprietary components, each with unique programming abstractions and interfaces. To facilitate straightforward deployment of edge applications, we introduce \gls{OSS} based \gls{RAN} on \gls{OTA} commercial spectrum with \gls{DevOps} capabilities. \gls{OSS} allows software modifications and integrations of the system components, e.g., \gls{EPC} and edge hosts running applications, required for new data pipelines and optimizations not addressed in standardization. Such an \gls{OSS} infrastructure enables further research and prototyping of novel end-user applications in an environment familiar to software engineers without telecommunications background. We evaluated the presented infrastructure with \gls{E2E} \gls{OTA} testing, resulting in 7.5MB/s throughput and latency of 21ms, which shows that the presented infrastructure provides low latency for edge applications.
\end{abstract}

\section{Introduction}

Edge computing is a prominent part of current and future cellular networks. It provides cloud resources, e.g., application-specific virtualization of computation and data, in the proximity of end-users. Virtualization at the edge decouples software from hardware, e.g., based on containers, that enables a variety of software-based deployment and coordination scenarios atop common hardware. As a coherent system, this promises application performance with more bandwidth and reduced latency for \glspl{UE} accessing the edge resources, while improving controls on privacy.

The ETSI \gls{MEC} standards aim to deliver these promises for the local \gls{RAN} atop \gls{5G} networks, allowing effective use of the up-and-coming \gls{5G} radio features. At the same time, \gls{SDN} and \gls{NFV} coupled \glspl{RAN} start to resemble edge platforms, as virtualization facilitates \glspl{MNO} building \gls{RAN} features on general-purpose hardware. This can address several scaling problems of \glspl{RAN} \cite{li2015software}, including dynamic network resource allocation, and network hardware ossification.

However, edge application development and deployment are currently challenging in \glspl{RAN}. First, \glspl{RAN}, with or without \gls{MEC}, have challenging requirements on \gls{E2E} latency, energy consumption and reliability with both up and downlinks \cite{condoluci2016enabling, mezzavilla2018end}. Second, edge applications are envisioned as both data- and computation-intensive. Thus, application deployment is not straightforward concerning the capabilities of edge hosts, application requirements, and environmental parameters of the \gls{RAN}. Only the consideration of all these aspects makes envisioned zero-latency applications possible. These issues are aggregated with multiple parallel applications on the same infrastructure.

Further, the current \gls{MEC} standardization offers less \gls{API} endpoints for software development than existing cloud computing environments. Therefore, linking development and actual deployment may be hard for end-user application developers, more familiar with current cloud environments. Coincidentally, \glspl{MNO} may be unable to introduce the required \glspl{API} for developers, for the \gls{RAN} components being proprietary. However, if the \gls{RAN} would be based on \gls{OSS}, then \gls{MNO} could create the required bridging interfaces. Using existing mature \gls{OSS} development platforms, which support the current mainstream software engineering processes, the challenges mentioned above could be thus addressed. In this context, we present an \gls{OSS} infrastructure for software-based \gls{5G} \gls{RAN} development, where \gls{DevOps} simplifies the deployment of \gls{E2E} \gls{OTA} applications to the edge, that also satisfies the attributes of a micro-operator, as discussed in \cite{ahokangas2016future}.

The paper is organized as follows. Section \ref{ch:background} discusses the related paradigms in software and telecommunications research. Section \ref{ch:infra} details the infrastructure design. Section \ref{ch:eval} evaluates the proposed infrastructure to deploy the network \gls{OTA} with \gls{RAN} hardware and spectrum licenses provided by the University of Oulu. Discussion follows in Section \ref{ch:discussion}, and lastly the contribution is concluded in Section \ref{ch:concl}.

\section{Background}\label{ch:background}

IoT and edge computing necessitate application development for a distributed environment, where virtualization provides a deployment mechanism. In \gls{5G}, virtualization is realized through \gls{SDN} and \gls{NFV} technologies. Combined, these technologies pave the way for the microservice architectural model \cite{balalaie2015migrating}, which can with \gls{DevOps} practices speed up software development for the edge.

\subsection{Virtualization}\label{virtualization}

\gls{SDN} and \gls{NFV} use virtualization to decouple application control and data into separate planes and to provide programmability and flexibility to network management through software. This approach supports adjusting the edge infrastructure and the network to user demands while considering the locally available resources. 

To realize virtualization on general-purpose hardware, containers have been reported to produce less operational overhead and better performance \cite{amaral2015performance, mao2015minimizing} than virtual machines that are based on virtualized operating system layer instead of hardware layer \cite{larisch2018alto}.

From a developer's perspective, managing containers means managing applications rather than machines. This enables the same deployment environment reducing inconsistencies between development and production, which improves deployment reliability and speed.

For \glspl{MNO}, containers provide an isolation mechanism which avoids edge applications from interfering with the host computer's, or other edge applications' execution. As the host computers solely manage containers, limiting operating system versions on edge hosts is easier, making the infrastructure more maintainable \cite{burns2016borg}. Container deployments spanning many physical host computers can then be managed with an orchestrator, which automates container creation, removal, restoration, and updating without interfering with the availability of the \gls{RAN} or the end-user edge applications.

\subsection{Microservices and -operators}

Microservices offer a distributed, cloud-native architectural model, which refines and simplifies service-oriented architecture \cite{balalaie2015migrating, amaral2015performance}. For example, microservices follow the share-nothing philosophy, to support agile business methods and to promote software isolation and autonomy \cite{richards2015microservices}. Microservices may be based on various programming languages, middleware and data stores. Each service runs in isolation while communicating through lightweight \glspl{API}. The microservice model is seen to provide isolation and scalability elastically and platform-agnostically. However, stable functional decomposition of the application into services is required with supporting system services, e.g., for service discovery, system configuration, load balancing, and fault detection.

In general, microservice development requires knowledge of distributed systems deployment, e.g., the tradeoffs between communication and computation overheads to realize the expected benefits \cite{balalaie2015migrating, dmitry2014micro}. This is further challenging with \gls{MEC}, which generally aims to optimize the connection between \glspl{UE} and \gls{MEC} services. 

From the \glspl{MNO} perspective, predicting the required computational and communication capabilities is challenging for a single site of operation in a traditional \gls{RAN}. A distributed \gls{MEC} platform adds extra variables into the challenge, such as application-specific resource migration within and between sites. Therefore, facilities for testing, deploying, and monitoring microservices becomes crucial in a cellular network due to multiple sites with varying edge computing capabilities.

In dense cellular deployments, \glspl{MNO} also face challenges in allocating radio resources with limited spectrum, managing interference, and in load-balancing cells. Local geographical \glspl{MNO} may thus be needed to manage wireless connectivity for \glspl{UE} effectively while maximizing throughput, latency, connectivity, or any other goal \cite{gudipati2013softran, xu2011cellular, wang2011untold}. Further, \glspl{MNO}' infrastructure is challenged by ever-increasing throughput and latency requirements of the \gls{5G} applications. If resources on wider area networks need to be accessed, latencies will grow, thus on-site data centers are a viable choice for edge applications on \gls{5G}. As such, micro-operators could be the first \glspl{MNO} relying solely on \gls{SDN} and \gls{NFV} principles, due to lack of legacy hardware and legacy application requirements from previous generations of cellular connectivity. Combined with agile business perspectives of microservices, such as the ability to cut long-term commitment and bottlenecks by realizing software as a set of complete, cohesive, and coupled services, micro-operators could thus provide an edge platform with customizability, low latency, and high reliability, that is unrealistic for traditional \glspl{MNO}. 

\subsection{DevOps}

\gls{DevOps} paradigm provides good software quality and frequent delivery mechanisms, for development teams of any size, facilitating enhanced value creation for end-users \cite{killalea2016hidden}. \gls{DevOps} allows software platforms to be offered as a \gls{PaaS} for \gls{E2E} application development. \gls{MEC} applications could be independently developed and tested without the intervention of the \gls{MNO}.

\gls{DevOps} utilizes \gls{CD}, \gls{CI}, and \gls{CM}. \gls{CD} and \gls{CI} enable on-demand deployment of software through automated mechanisms \cite{amaral2015performance}. When the number of deployments increases, as with microservices, \gls{CD} and \gls{CI} become essential for frequent software releases. \gls{CM} then provides microservice developers and the \gls{MNO} with performance-related feedback of the releases, e.g., to detect operational anomalies, and profiling.

\begin{figure*}[!t]
    \includegraphics[width=0.9\textwidth]{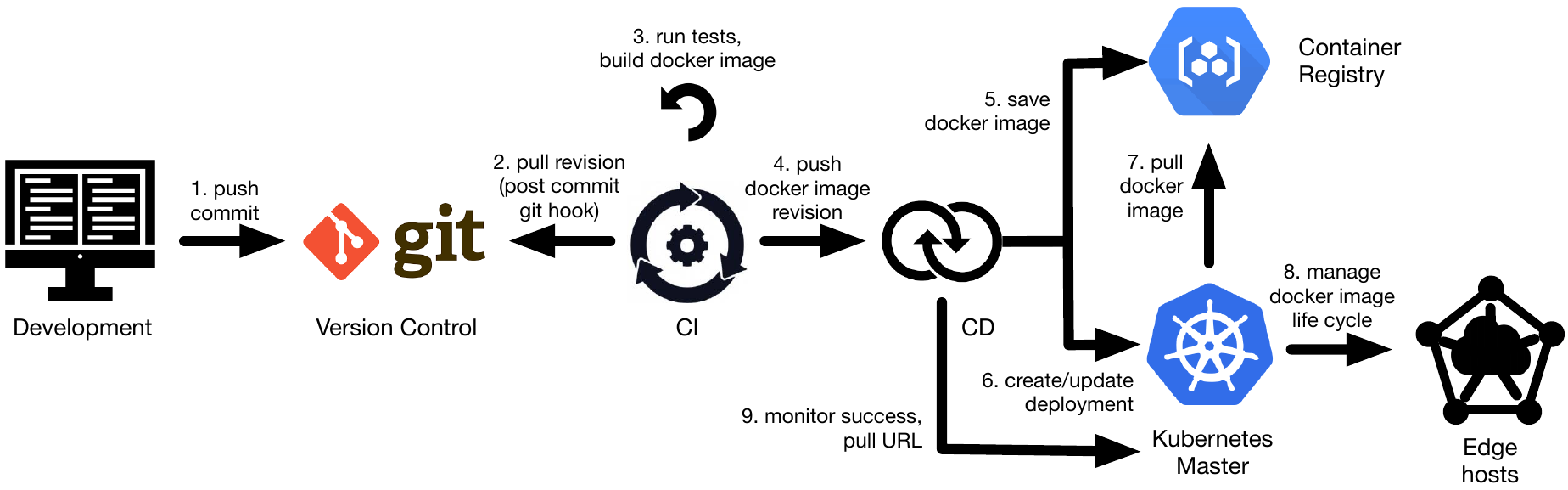}
  \centering
  \caption{Workflow and components of the DevOps platform.}
  \label{fig:paas}
\end{figure*}

\subsection{Related work}\label{ch:sota}

Virtualized cellular networks without \gls{OTA} testing have been studied widely \cite{mao2015minimizing, vallati2015exploiting, peng2016fog, elkhatib2017using, hsieh2018mobile, munoz2018integration, amemiya2018site, bolivar2018deployment, walia20175g, siriwardhana2018micro, tseng2018micro}. \gls{OTA} testbeds without \gls{MEC} have been installed on LTE \cite{kamarainen2017measurement, eckermann2018tinylte, hadvzic2017edge, li2018mobile} and, to avoid regulated spectrum, on WiFi \cite{streiffer2017eprivateeye, carmo2018slicing}. Proprietary \gls{OTA} testbeds with \gls{MEC} exist \cite{cattaneo2018deploying, piri20165gtn, mangiante2017vr, karimi2017evaluating}, but without consideration of \gls{DevOps} practices. In general, \gls{DevOps} practices have rarely been installed on \gls{LTE}, but one study did provide \gls{IaaS} services \cite{gosain2016enabling} making such deployments possible, and another one has offered \gls{DevOps} \cite{van2017mec} on WiFi. As such, our study essentially combines the \gls{OTA} environment of \cite{gosain2016enabling}, but replaces \gls{IaaS} as in \cite{van2017mec} with \gls{PaaS}, thus relieving the developer to only maintain an application instead of a server, as described in §\ref{virtualization}. To the best of our knowledge the research in this niche is yet uncharted.

\section{Infrastructure design}
\label{ch:infra}

Testbeds are a requisite for experimental research as developed new technologies, methods, and features can be validated on-site. On the one hand, \gls{OSS} testbeds are preferred for flexibility and openness, but they tend to be inferior in performance and may lack some capabilities in comparison with their proprietary counterparts. On the other hand, the business goals of commercial testbeds may differ from research goals and focus on, e.g., standardization or off-the-shelf components. As an example, current off-the-shelf \gls{SDR} hardware allows building cellular deployments on virtualization and \gls{OSS} \cite{eckermann2018tinylte}, but the performance and features are underdeveloped.

In the corporate market, Kubernetes has emerged as the leading solution for development deployments for the edge \cite{efrati2017cloud, burns2016borg}. Moreover, microservice infrastructures have been promoted by major cloud computing infrastructure providers, such as Google, Microsoft, and Alibaba, with varying capabilities for management.

As we set the requirement to use the same infrastructure in both development and deployment, we built our platform with Kubernetes. In this regard, Kubernetes provides built-in isolation, prioritization, and scheduling mechanisms for separating developer applications from the actual deployments on the same infrastructure. Kubernetes supports Docker for container-based application deployment, which we used for microservice development in the proposed infrastructure. Docker provides lightweight virtualization for distributed edge application components; thus widespread deployments become possible with minimized hardware requirements.

Fig. \ref{fig:paas} presents an overview of the proposed workflow, where \gls{DevOps} services are installed on \gls{MEC} hosts on Kubernetes. Thanks to industry adoption of Kubernetes in software engineering, open-source \gls{DevOps} services like revision control (e.g., Gitea), \gls{CI} and \gls{CD} (e.g., Drone), and \gls{CM} (e.g., Prometheus) are easy to integrate. By installing these \gls{DevOps} tools on Kubernetes we base the initial configuration of the testbed on \gls{OSS}, avoiding dependency on a single tool provider. Should a \gls{DevOps} component prove to be unsuitable for our needs, we can replace it more cost-effectively than a proprietary cloud-based service. As a result, the unity of these components contains most of the functionality specified in the \gls{MEC} reference architecture \cite{etsimec003} (Fig. 6-1). The reference architecture is further described in \cite{taleb2017multi} (§V.B), with which we can translate the ETSI standard to Kubernetes terminology: APP is a deployment in Kubernetes, VIM is a container registry, CFS is kubectl interfacing through Web UI, and LCM is kubectl interfacing through JSON API.

New \gls{E2E} edge applications are deployed as follows (see Fig. \ref{fig:paas}): a developer places application source-code in a git-repository. Once the developer pushes changes to the master branch of the repository (1), an automatic post-commit git hook pulls the source-code to the \gls{CI} server (2), which runs the developer-specified tests (3), and if successful, attempts to build the application using its Dockerfile. If successful, the \gls{CI} creates a Docker image, passes it to \gls{CD} (4), which saves it (5), and commands Kubernetes to deploy the application using the built Docker image (6, 7). If Kubernetes is able to provision the application (8), the \gls{CD} server returns the application's DNS address from Kubernetes (9), which is then passed onto the developer to be used to access the application from an \gls{UE}.

\begin{figure*}[ht]
    \centering
    \begin{minipage}[b]{0.45\textwidth}
        \includegraphics[width=\textwidth]{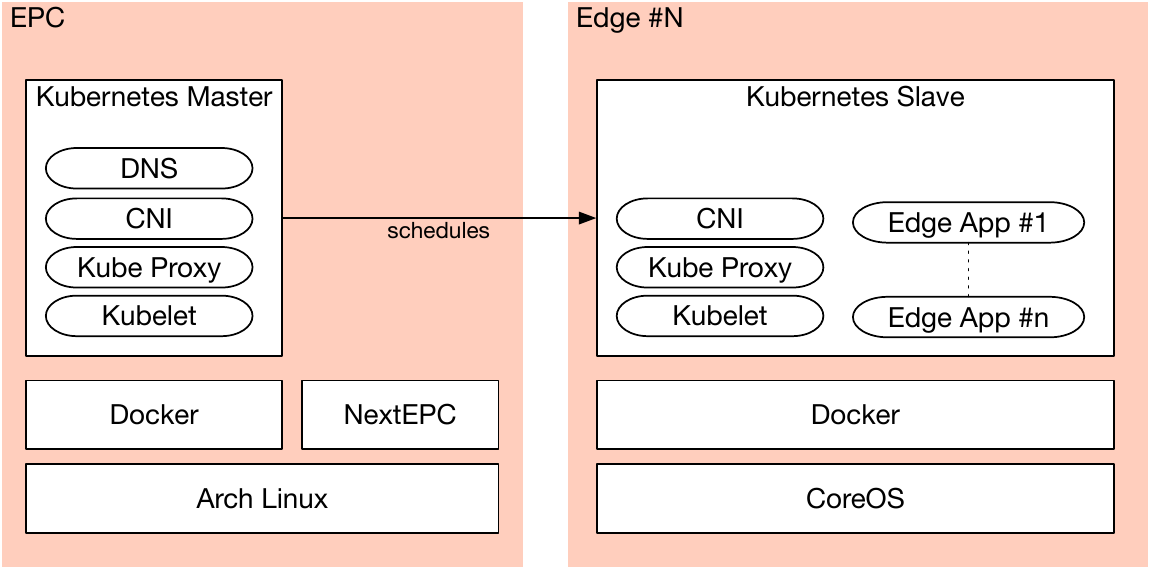}
        \caption{Software stacks of the edge hosts. Left: the primary host running Kubernetes master and the \gls{EPC}. Right: the edge hosts hosting the edge applications.}
        \label{fig:softwarestack}
    \end{minipage}
    \quad \quad 
    \begin{minipage}[b]{0.45\textwidth}
        \includegraphics[width=\textwidth]{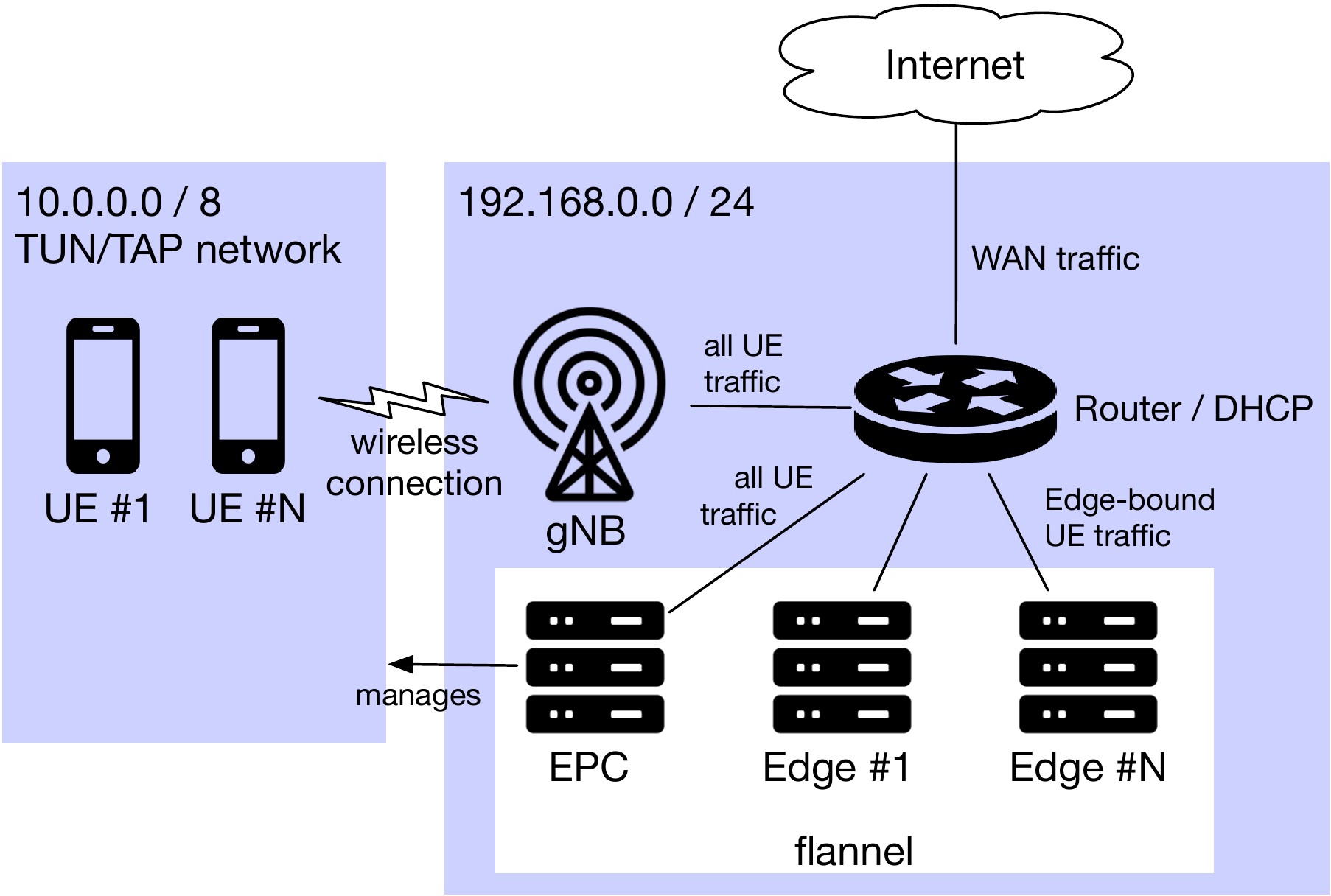}
        \caption{Platform network architecture.}
        \label{fig:network}
    \end{minipage}
\end{figure*}

Fig. \ref{fig:softwarestack} shows the software stacks of the primary host running Kubernetes master and the \gls{EPC}, and the edge hosts meant solely for hosting the edge applications. The primary host is built on Arch Linux, as it provides straightforward installation of a real-time Linux kernel. The edge hosts operate on CoreOS, that runs containers natively, and be used to form a cluster of bare-metal servers. The clusterization allows the CoreOS hosts to update the container runtime and kernel components automatically and in succession, resulting in zero-downtime operating system upgrades \cite{website:coreos}. Each physical host then runs the standard Kubernetes architecture, e.g., flannel as a CNI (Container Network Interface) for establishing a virtual network that attaches IPs for containers, Kube Proxy for routing between pods, and Kubelet as orchestration agent responsible for the control interface.

\subsection{Network architecture}

Fig. \ref{fig:network} shows the network architecture of our infrastructure. \glspl{UE} are assigned an IP from the \verb|tun/tap| device of the \gls{EPC}. As a part of the Kubernetes cluster, the \gls{EPC} can route \glspl{UE} to access the rest of the Kubernetes deployment, on which the load-balanced edge applications can run in a single network hop from the \gls{EPC}, facilitating guarantees on low latency.

The \gls{EPC} implementation must support low latency edge applications on the \glspl{RAN}. Thus a real-time Linux kernel is a requirement to optimize its network stack \cite{mao2015minimizing}. As discussed in \cite{panda2016netbricks}, \gls{EPC} should be left unvirtualized for \gls{NFV} management. Further, we installed kernel passthrough for the \gls{NIC} \cite{mao2015minimizing, panda2016netbricks}. As such, the \gls{EPC} is a scalable software component instead of a hard to upgrade proprietary \gls{DSP} hardware.
 
In this respect, we reviewed the current \gls{OSS} \gls{EPC} projects, as shown in Table \ref{tab:EPC}. We chose NextEPC as our environment for its ease of installation and support for \gls{3GPP} Release 13. Besides having the most up-to-date \gls{3GPP} conformance, NextEPC is also the only \gls{EPC} installable from a package manager and with an easy-to-use web-based subscriber management interface. 

\begin{table}[!ht]
\centering
\caption{Currently available OSS EPC packages.}
    \begin{tabular}{ l l p{3.5cm}}
    \toprule
    \textbf{Project} & \textbf{Language} & \textbf{Anecdotal summary} \\
    \midrule
    \href{https://github.com/acetcom/nextepc}{NextEPC} & C & \gls{3GPP} Release 13 support \\
    \midrule
    \href{https://github.com/mitshell/corenet}{corenet} & Python & Minimal 3G and \gls{LTE} \gls{EPC} \\
    \midrule
    \href{https://github.com/networkedsystemsIITB/NFV_LTE_EPC}{vEPC} & C++ & For performance testing \cite{jain2016comparison} \\
    \midrule
    \href{https://github.com/OPENAIRINTERFACE/openair-cn}{openair-cn} & C & \gls{3GPP} Release 10 support \cite{website:oai} \\
    \midrule
    \href{https://github.com/srsLTE/srsLTE}{srsLTE} & C++ & \gls{3GPP} Release 8 support \\
    \midrule
    \href{http://openlte.sourceforge.net/}{openLTE} & C++ & Elegant, but incomplete \cite{gomez2016srslte} \\
    \bottomrule
    \end{tabular}
\label{tab:EPC}
\end{table}

\section{Evaluation and results}\label{ch:eval}

We evaluated the proposed infrastructure with regard to communication latency between one \gls{UE} and the \gls{EPC} as infrastructure end-point, that is the focus of the edge paradigm. The evaluations are conducted \gls{OTA} using Nokia picocell as a base station, set at 10Mhz on band 7, using spectrum licenses of the University of Oulu. The \gls{EPC} is installed on commodity hardware running i3-6100. The results establish a baseline on which to set the performance expectations with application-specific virtualization.

\begin{figure}
  \centering
  \includegraphics[width=0.45\textwidth]{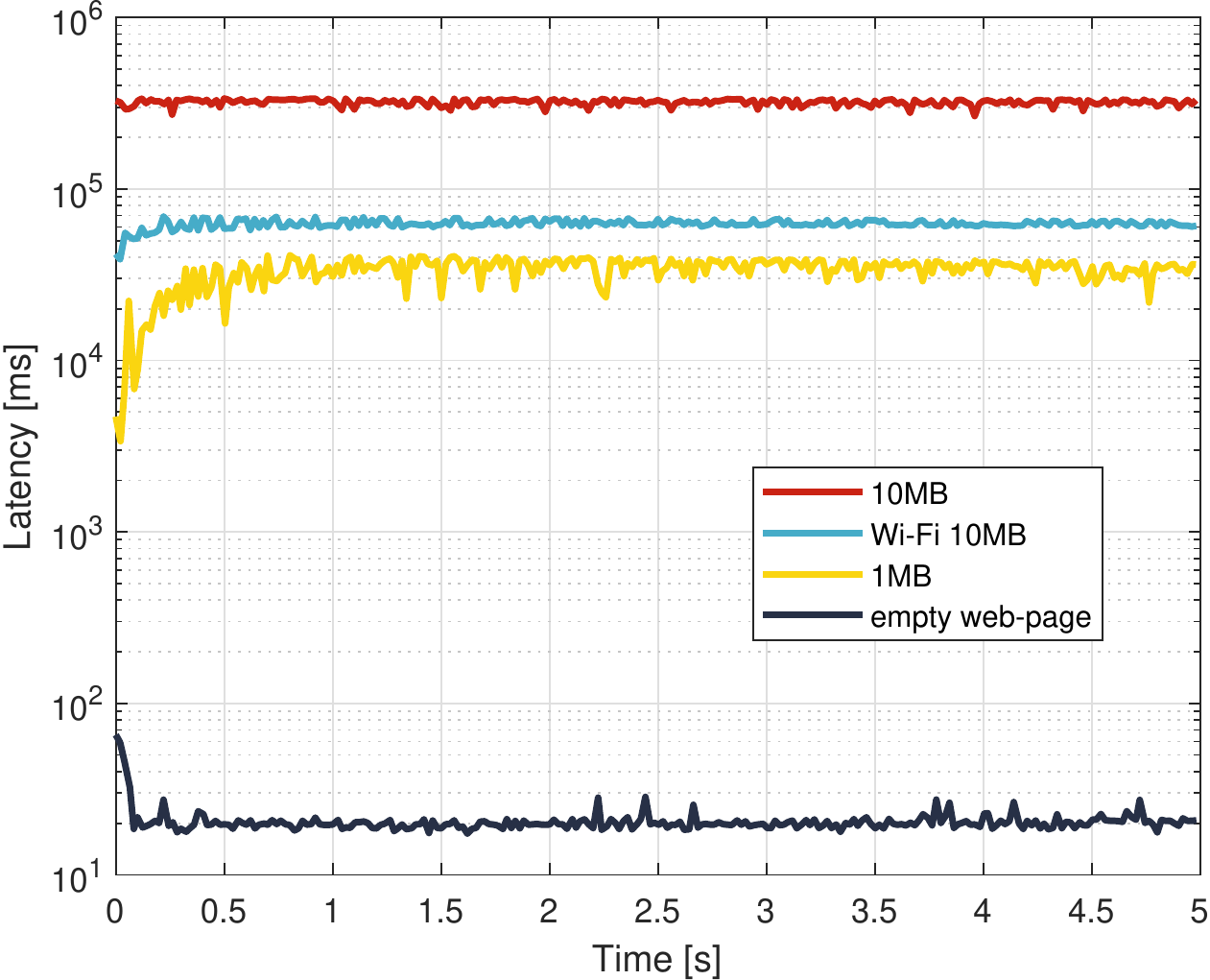}
  \caption{End-to-end latencies on four different data sets.}
  \label{fig:webpage}
\end{figure}

We measured latencies in four data sets, as shown in Table \ref{tab:testresults} and Fig. \ref{fig:webpage}. The first set served an empty web page. The rest of the tests served pre-generated binary blogs generated by \verb|/dev/rand|. The second test served a single 1MB binary file. The third test served a single 10MB binary file, and the fourth served 10MB binary file over WiFi as a baseline. All the tests sent 250 requests over 5 seconds in chunks of 50 requests per second. We used HTTP-based Vegeta\footnote{https://github.com/tsenart/vegeta} for benchmarking. The \gls{UE} was OnePlus 5T phone tethered to a MacBook via USB, which the MacBook used as a modem. 

The detailed datasets, software versions, and hardware component specifications are available on Github\footnote{https://github.com/toldjuuso/haavisto2019infra}. Given the results, further research is inevitable to debug anomalies such as the first request latency of an empty webpage of 65.6ms as shown in Fig. \ref{fig:webpage}. While WiFi does yield better performance, it is worth noting that \gls{LTE} has wider operation range thus is more suitable for applications requiring high mobility. Further, we note the \gls{EPC} uses userland implementations of the GPRS Tunneling Protocol (GTP), and Stream Control Transmission Protocol (SCTP), which are the transport protocols \gls{LTE} communication is based on. As both protocols have Linux kernel implementations, the source code of the \gls{EPC} could be modified to use the more privileged versions thus facilitating higher performance and lower latency. Moreover, integrating Data Plane Development Kit bindings would provide further network performance gains \cite{mao2015minimizing}.

\begin{table}[!ht]
\centering
\caption{Latency and throughput}
    \begin{tabular}{p{3.5cm} r r r}
    \toprule
    \textbf{Test} & \textbf{Mean} & \textbf{99th} & \textbf{Throughput} \\ 
    \midrule
    empty HTML document & 20.6ms & 45.9ms & 9.38MB/s \\
    \midrule
    1MB blob & 33.6s & 40.5s & 6.72MB/s \\
    \midrule
    10MB blob & 332.5s & 336.0s & 7.52MB/s \\
    \midrule
    10MB blob on WiFi & 62.5s & 68.7s & 38.65MB/s \\
    \bottomrule    
    \end{tabular}
\label{tab:testresults}
\end{table}

\section{Discussion}\label{ch:discussion}

In this paper, we have presented an \gls{OSS}-based infrastructure for development and deployment of distributed \gls{OTA} applications for \gls{MEC} atop \gls{5G} networks. In essence, we provide low latency computation resources for end-user applications. In comparison with previous work, we present the first \gls{RAN} infrastructure combining \gls{OTA}, \gls{DevOps} and \gls{PaaS}. We observed that the network can handle the latency of 21ms and bandwidth of 7.5MB/s in \gls{E2E} \gls{OTA} testing. Introducing \gls{DevOps} methodology can lower the learning curve to create \gls{5G} \gls{MEC} services for end-user application developers. Moreover, the \gls{OSS} architecture allows continuation research of off-the-spec \gls{RAN} modifications which may be next to impossible to implement in proprietary environments.

The contribution stitches together interfaces of telecommunications and computer science through software engineering. As a result, we formed a mobile network operator, which correspondingly provides edge computing in a similar manner as industry \gls{PaaS} businesses offer cloud computing. The authors are not aware of any similar pre-existing counterparts being presented in literature, and hence the research is novel. We continue to conduct practical research in the presented architecture. Thus, this pragmatic instance helps to envision the role of future mobile network operators while retaining a level of lucidness in research through pragmatism.

Regarding virtualization, the current architecture requires hardware to support Docker to act as a schedulable host. Another possibility would be to use Wasm \cite{haas2017bringing} as the application development environment, with reduced communication and computational overhead stipulating virtualization change from Docker to Google V8. V8 supports more processor architectures than Docker, which would enable a broader range of computational units to join the distributed platform.

The presented hardware platform could also be improved, targeting enabling implementations of deadline-critical \gls{NFV} functionality using open-source hardware such as Power9 architecture and Open Compute Project network hardware. These implementations could reside on more profound levels of software abstraction, thus resulting in better performance and topics of future research.

Finally, the University of Oulu owns the \gls{RAN} hardware, and has the regulated spectrum licenses to deploy the network \gls{OTA}. This enables a future scenario with eSIMs, in which the campus area has an open cellular network for students and faculty members. In such scenario, subscribers could programmatically have SIM credentials created using QR codes on the campus hallways while aspiring developers could deploy software both as \gls{RAN} infrastructure components, but also as end-user \gls{E2E} applications accessible over-the-air within the campus periphery.

\section{Conclusion}\label{ch:concl}

An infrastructure for \gls{RAN} was presented, where the target is to enable low latency connections for \glspl{UE} in the context of \gls{5G} networks for \gls{MEC}. Our evaluation shows that the network, at the premises of the University of Oulu, can handle latency of 21ms and bandwidth of 7.5MB/s in \gls{E2E} \gls{OTA} testing. These results provide a real-world baseline for further improving such infrastructures towards the goals of \gls{5G} and edge computing. The \gls{OSS} infrastructure allows off-the-spec \gls{RAN} modifications, this is difficult to realize in proprietary environments. Our future work addresses improvements concerning the \gls{EPC} and edge host hardware. Moreover, we envision further real-world \gls{OTA} application deployments and studies with industry developers, researchers, and students on the presented open cellular network. 

\section{Acknowledgements}\label{ch:ack}
This research is financially supported by Academy of Finland 6Genesis Flagship (grant 318927) and by the AI Enhanced Mobile Edge Computing project, funded by the Future Makers program of Jane and Aatos Erkko Foundation and Technology Industries of Finland Centennial Foundation.

\bibliographystyle{ieeetr}
\bibliography{main}
\end{document}